\documentstyle[12pt]{article}
\def\beq{\begin{equation}}
\def\eeq{\end{equation}}
\def\bea{\begin{eqnarray}}
\def\eea{\end{eqnarray}}
\def\bq{\begin{quote}}
\def\eq{\end{quote}}

\def\NP{{\it Nucl.Phys.} }
\def\PL{{\it Phys.Lett.} }

\def\ZP{{\it Z.Phys.} }

\def\gappeq{\mathrel{\rlap {\raise.5ex\hbox{$>$}}
{\lower.5ex\hbox{$\sim$}}}}

\def\lappeq{\mathrel{\rlap{\raise.5ex\hbox{$<$}}
{\lower.5ex\hbox{$\sim$}}}}

\begin{document}
\pagestyle{empty}
\vspace*{5mm}
\begin{center}
{\bf   NON-EQUILIBRIUM TIME EVOLUTION}\\
{\bf  IN QUANTUM FIELD THEORY} \\
\vspace*{1cm} 
{\bf Christof WETTERICH}$^{*)}$ \\
\vspace{0.3cm}
Institut f\"ur Theoretische Physik \\
Universit\"at Heidelberg \\
Philosophenweg 16, D 69120 Heidelberg \\
\vspace*{0.3cm}
{\tt C.Wetterich@thphys.uni-heidelberg.de}\\
\vspace*{2cm}  
{\bf ABSTRACT} \\ \end{center}
\vspace*{5mm}
\noindent
The time development of equal-time correlation functions in
quantum mechanics and quantum field theory is described by an
exact evolution equation for generating functionals. This permits
a comparison between classical and quantum evolution in
non-equilibrium systems.
\vspace*{6cm}

\noindent 
\rule[.1in]{10cm}{.002in}

\noindent
$^{*)}$This work was performed in part at CERN, Geneva.
\vspace*{0.5cm}

\vfill\eject

\setcounter{page}{1}
\pagestyle{plain}

In a statistical description of nature only expectation values 
or correlations are observable. The dynamical laws describe how a given
(complete) set of correlation functions at some initial time $t_0$ has
evolved at some later time $t$. In a very general context this constitutes
a
differential evolution equation for the correlation functions or their
generating functionals. Different dynamical laws, i.e. the difference
between a classical and a quantum description, manifest themselves 
only in different evolution equations. For quantum mechanics of a few
degrees
of freedom we are accustomed to use the Schr\"odinger equation for pure
states and to evaluate time-dependent expectation values in an ensemble
specified by a density matrix. The classical counterpart are Newton's
equations with the ensemble described by a probability distribution for
initial values. For many degrees of freedom a more direct formulation of
the
dynamics in terms of an evolution equation for generating functionals of
correlation functions may be advantageous. Then one only treats with the
relevant observable information and may avoid additional complication in
intermediate steps of their computation. For classical equations of motion
such an evolution equation has been established recently \cite{W1}. It is
the purpose of this note to develop the counterpart for quantum mechanics
and
quantum field theory.

The resulting evolution equation in quantum field theory is a functional
differential equation. Its practical use depends on the ability to find
realistic truncations for the time-dependent effective action - the
generating functional for the 1PI-Green functions. Nevertheless, already
at
the present formal stage a comparison between classical and quantum
evolution
equations sheds some light on current dynamical simulations of
non-equilibrium quantum field theory problems within a classical
approximation for the field equations. It also provides a systematic
framework to compute quantum corrections to a classical evolution, as
relevant, for example, for inflationary periods in early cosmology.

Consider a system with an arbitrary number of degrees of freedom described
by
conjugate operators $Q_i, P_i, i = 1,\ldots n$,  $[Q_i,P_i] =
i\delta_{ij},
[Q_i,Q_j] = [P_i,P_j] = 0,~~~(\hbar \equiv 1)$. The
Hamiltonian\footnote{Systems with different masses $m_i$ can be brought to
this form by simultaneous rescaling of $Q_i$ and $P_i$ - in fact, we could
also use a scaling with $m = 1$. Our formalism can be generalized for
arbitrary
$H[P,Q]$ but we restrict the discussion here to the most common form
quadratic in the momenta.} is given by
\beq
H = {1\over 2m}~P^2 + V[Q]~,
\label{one}
\eeq
where $P^2 = P_i~P_i$ is a scalar product (summation of repeated indices
is always implied) and $V[Q] = V(Q_1\ldots Q_n)$. In the Heisenberg
picture
the operators $Q$ and $P$ depend on time according to
\beq
\dot Q_i = {1\over m}~P_i~,~~~\dot P_i = -V_i[Q]
\label{two}
\eeq
(We use a notation $V_i = {\partial V\over \partial Q_i}, V_{ij} =
{\partial^2 V \over \partial Q_i \partial Q_j}$, etc.). Let us now define
the
generating functional
\beq
Z[j,  h, t] = {\rm Tr} (e^{jQ(t)+hP(t)}~\rho )
\label{three}
\eeq
where $\rho$ is the density matrix which is time-independent in the
Heisenberg picture. It is a function (or functional for
$n\rightarrow\infty$)
of the time-independent sources $j_i$ and $h_i$. The coefficients of its
Taylor expansion
\beq
Z[j,h,t] = \sum^\infty_{k=0}~~ \sum^\infty_{\ell =0}~~{1\over k!\ell
!}~~z^{(k,\ell )}_{q_1\ldots q_k,r_1\ldots r_\ell}(t) j_{q_1}\ldots
j_{q_k} ~~
h_{r_1}\ldots h_{r_\ell}
\label{four}
\eeq
are time-dependent symmetrized expectation values of correlation
functions.
For a one-component example $(n=1)$
\beq
z^{(k,\ell )}(t) = \langle (Q^k P^\ell )_s(t)  \rangle = 
{\rm Tr} \left\{(Q^k(t) P^\ell (t))_s\rho\right\}
\label{five}
\eeq
the symmetrized ordering $(~~~~)_s$ stands for an equally weighted sum
over
all ${(k+\ell)!\over k!\ell!}$ possibilities to form different chains of
$k$
operators $Q$ and $\ell$ operators $P$, i.e., 
\beq
(Q^2P^2)_s = {1\over 6} (Q^2P^2 + QP^2Q + P^2Q^2 + PQ^2P + QPQP + PQPQ)
\label{six}
\eeq
The generalization to arbitrary $n$ is obvious. Knowledge of $Z[j,h,t]$
contains the complete information about the system. Macroscopic quantities
and thermodynamic functions can be expressed in terms of correlation
functions \cite{BB}.

The time-dependence of $Z$ obeys the evolution equation ($\partial_t Z$ is
the time derivative at fixed $j$ and $h$)
\beq
\partial_t Z = i {\rm Tr}  \left([H, e^{jQ+hP}]~~\rho\right)
\label{seven}
\eeq
Our aim is to express the right-hand side in terms of $Z$ and its
derivatives
with respect to $j$ and $h$. In order to deal with the problem of ordering
the non-commuting operators we use
\beq
i[H,e^{jQ+hP}] = {1\over m}~j_i ~{\partial\over\partial h_i}~e^{jQ+hP} +
iV\left[{\partial\over\partial\beta}\right]~\left[e^{\beta Q},
e^{jQ+hP}\right]\vert_{\beta = 0}
\label{eight}
\eeq
and evaluate the commutator with the Campbell-Barker-Hausdorff formula
\beq
\left[ e^{\beta Q}, e^{jQ+hP}\right] = \left(e^{{i\over 2}\beta h}
-e^{-{i\over 2}\beta h}\right) e^{(\beta+j) Q+jP}
\label{nine}
\eeq
For each of the two contributions on the right-hand side, one can express
${\partial\over\partial\beta}$ by an appropriate combination of 
${\partial\over\partial j}$ and $h$. This leads to
\beq
i\left[V[Q], e^{jQ+hP}\right] = i\left(V\left[{\partial\over\partial j} +
{i\over 2}  h \right] -V\left[{\partial\over\partial j} -
{i\over 2}  h \right]\right) e^{jQ+hP}
\label{ten}
\eeq
and (\ref{eight}), (\ref{ten}) can now be inserted in the trace
(\ref{seven}). Our final result is a linear partial differential equation
for
the evolution of $Z$ with a simple structure
\bea
\partial_t Z &=& {\cal L}^{(Z)} Z \nonumber \\
{\cal L}^{(Z)} &=& {1\over m}~j_i~{\partial\over\partial h_i} + i
\left( V\left[{\partial\over\partial j} + {i\over 2}~h\right] 
-V \left[{\partial\over\partial j} - {i\over 2}~h\right]\right)
\label{eleven}
\eea

For a comparison with the time evolution of $Z$ in classical statistics 
\cite{W1} we expand $V$ in powers of $h$
\bea
&&iV\left[{\partial\over\partial j} + {i\over 2} h\right] - iV
\left[{\partial\over\partial j} - {i\over 2} h\right] = -V_i 
\left[{\partial\over\partial j}\right] h_i
+ {1\over 24} V_{ijk}~\left[{\partial\over\partial j}\right] 
h_i h_j h_k \nonumber \\
&&-{1\over 1920} V_{ijk\ell m} 
\left[{\partial\over\partial j}\right] 
h_i h_j h_k h_\ell h_m +\ldots \hfill
\label{twelve}
\eea
The first term in this expansion reproduces the classical evolution
equation.
This leads to the important observation that for linear equations of
motion
($H$ quadratic in $Q$ and $P$) there is no difference in the time
evolution
of correlation functions between quantum statistics and classical
statistics!
The higher order terms appearing for non-linear equations of motion can be
viewed as quantum corrections to the classical evolution. Restoring
$\hbar$,
they involve powers $\hbar^2$, $\hbar^4$, etc. For $V$ containing only up
to
quartic terms the quantum correction $\Delta {\cal L}^{(Z)}_{QM}$ reads
explicitly $(V_{ijk} \equiv V_{ijk} [0])$
\bea
{\cal L}^{(Z)} &=& {\cal L}^{(Z)}_{cl} + \Delta {\cal L}^{(Z)}_{QM}
\nonumber \\
{\cal L}^{(Z)}_{cl} &=& {1\over m} ~j_i ~ {\partial\over\partial h_i} ~
-h_i~V_i \left[{\partial\over\partial j} \right] \nonumber \\
\Delta {\cal L}^{(Z)}_{QM} &=& {1\over 24} ~V_{ijk}~~h_i~h_j~h_k +
{1\over
24} ~V_{ijk\ell}~h_i~h_j~h_k~
{\partial\over\partial j_\ell}
\label{thirteen}
\eea

It may be instructive to consider two examples. We first take a single
anharmonic oscillator with
\beq
V(Q) = {1\over 2}~ m \omega^2 (Q^2 + {4\over 3} \sqrt{2 m\omega}\gamma Q^3
+
2 m\omega\delta Q^4
\label{fourteen}
\eeq
In terms of a complex source
\beq J = {1\over \sqrt{2m\omega}} (j+im\omega h)
\label{15}
\eeq
which is conjugate to the creation operator $a^\dagger$ of the harmonic
oscillator the evolution equation reads
\bea
\partial_t Z &=& -i\omega \left\{  J^*{\partial\over\partial J^*} - J
{\partial\over\partial J}+ (J^* - J) \left[\gamma
\left({\partial\over\partial J^*} +
{\partial\over\partial J} \right)^2 + \delta \left(
{\partial\over\partial J^*} + {\partial\over\partial J}\right)^3\right]
\right.\nonumber \\
&&\left.+ {\gamma\over 12} \left( J^* - J\right)^3 + {\delta\over 4}
\left(
J^* - J\right )^3
\left( {\partial\over\partial J^*} + {\partial\over\partial J} \right)
\right\} Z
\label{sixteen}
\eea
In our picture the state of the system at a given time $t$ is described by
$Z[J,t]$. For the harmonic oscillator $(\gamma = \delta = 0)$ the
stationary
states or fixed points of $Z~(\partial_t Z = 0)$ are exactly those for
which
all terms in $Z$ involve an equal number of powers of $J$ and $J^*$. They
correspond to incoherent mixtures of eigenstates of the number operator
$a^\dagger a$ or energy eigenstates. It is intriguing that due to the
equivalence of the quantum and classical evolution for linear equations of
motion the quantum mechanical energy eigenstates can also be viewed as
classically stationary probability distributions for commuting coordinate
and
momentum variables.
For $\gamma , \delta \not= 0$ these particular states are not stationary
any
more. We know, nevertheless, that Eq. (\ref{sixteen}) must admit an
infinite
number of fixed point solutions which correspond exactly to incoherent
mixtures of energy eigenstates of the anharmonic oscillator. This follows
>from the general observation that Eq. (\ref{seven}) is equivalent to
\beq
\partial_t Z = -i {\rm Tr} \left( e^{jQ+hP}[H,\rho]\right)
\label{seventeen}
\eeq
and $H(t) = H(t_0)$. Every $\rho$ which commutes with $H$ defines a
stationary state. An important qualitative difference between the quantum
and
classical evolution equation can be easily seen if we neglect $\delta$.
The
quantum mechanical evolution equation for $\ln Z$ contains a ``correction"
term ${\gamma\over 12}(J^*-J)^3$ which acts as an additional constant
``force" on the connected correlation function
for $P^3$, i.e., $\Delta \partial_t <P^3>_c = {1\over 12} \omega\gamma
(2m\omega )^{3/2}$. A cubic potential is not bounded from below. The local
minimum at the origin is separated from the unstable part by a barrier. A
classically stable initial ensemble with all energies below the barrier
becomes unstable in quantum mechanics due to tunneling, as reflected by
the
additional ``force". The essential features of this effect are not changed
if
we restore stability by a small positive non-zero $\delta$. Once
reexpressed
in terms of real sources $j$ and $h$, Eq. (\ref{sixteen}) is a real
partial
linear differential equation for a function of three variables $Z(j,h,t)$.
A
linear combination of two solutions is again a solution. A numerical
solution of this equation contains at once the information about the time
evolution of all expectation values of arbitrary powers of $Q$ and $P$! In
particular, it seems interesting to investigate the outcome for
classically
chaotic systems.

Our second example is a linear chain of oscillators (with mass $m$ scaled
to one and $d = 1$) 
\beq
H = \sum_i \left\{ {1\over 2} P^2_i + {1\over 2} \mu^2 Q^2_i + {1\over 3}
\nu a^{-d/2} Q^3_i + {1\over 8} \lambda a^{-d} Q^4_i + {1\over
2a^2}~(Q_{i+1}-Q_i)^2\right\}
\label{eighteen}
\eeq
Here $a$ is the distance between two oscillators on the chain and we take
periodic boundary conditions with a fixed length $\Omega = \sum^{n}_{i=1}
a$.
The evolution equation reads
\bea
\partial_t Z &=& \sum_i \left\{ j_i {\partial\over\partial h_i} + {1\over
a^2}
h_i \left( {\partial\over\partial j_{i+1}} - 2{\partial\over\partial j_i}
 + {\partial\over\partial j_{i-1}}\right)\right. \nonumber \\
&&- h_i \left( \mu^2 {\partial\over\partial j_i} + \nu a^{-d/2} \left(
{\partial\over\partial j_i}\right)^2 + {1\over 2} \lambda a^{-d} 
\left({\partial\over\partial j_i}\right)^3 \right) \nonumber \\
 && \left. + {1\over 12} \nu a^{-d/2} h^3_i + {1\over 8} \lambda a^{-d}
h^3_i 
{\partial\over\partial j_i} \right\} Z
\label{nineteen}
\eea
Again, the last two terms are the quantum corrections.

The transition to a field theory can be made by taking the limit
$a\rightarrow 0$ while keeping a non-zero ``volume" $\Omega$. With the
replacements
$a^{-d/2}Q_i\rightarrow \tilde\varphi (x), ~~a^{-d/2}P_i = \tilde\pi (x),
~~a^d
\sum_i\rightarrow\int d^dx, ~~ a^{-(d/2 +1)} (Q_{i+1}-Q_i)\rightarrow
\partial \tilde\varphi /\partial x_i$ -- we generalize here to an
arbitrary
dimension $d$ -- the Hamiltonian (\ref{eighteen}) becomes
\bea
H&=& \int d^d x \left\{ {1\over 2}~\tilde\pi^2(x) + {1\over 2} \partial_i
\tilde\varphi (x) \partial_i \tilde\varphi (x) + U(\tilde\varphi (x))
\right\}\nonumber \\
U(\tilde\varphi (x)) &=& {1\over 2} \mu^2 \tilde\varphi^2 (x) + {1\over 3}
\nu \tilde\varphi^3 (x) + {1\over 8} \lambda \tilde\varphi^4 (x)
\label{twenty}
\eea
This is the Hamiltonian of a relativistic scalar field theory with field
equations $\tilde\pi (x) = \dot{\tilde\varphi}(x)$, $\dot{\tilde\pi} =
\ddot{\tilde\varphi} = \Delta \tilde\varphi - \partial U/\partial
\tilde\varphi$. Since also the commutation relation
$(a^{-d}\delta_{ij}\rightarrow \delta (x - x^\prime))$
\beq
[\tilde\varphi (x) , \tilde\pi (x^\prime)] = i\delta (x-x^\prime)
\label{twentyone}
\eeq
is covariant under Lorentz transformations, our system describes a
Lorentz-invariant quantum field theory. Here $Z$ becomes a functional of
the
sources $j_\varphi(x), j_\pi(x)$ which are related 
to the sources in the discrete version by 
$a^{-d/2} j_i \rightarrow j_\varphi (x), ~~ a^{-d/2} h_i\rightarrow j_\pi
(x), ~~ a^{-d/2}{\partial\over\partial j_i} = {\delta\over \delta
j_\varphi
(x)}$ and 
$a^{-d/2}{\partial\over\partial h_i} = {\delta\over\delta j_\pi(x)}$,
i.e.,
$Z = {\rm Tr} \left\{ \exp (\int d^d x[j_\varphi(x) \tilde\varphi (x) +
j_\pi
(x) \tilde\pi (x) ])\rho \right\}$.
We finally arrive at the time evolution equation for a scalar quantum
field
theory
\bea
\partial_t Z[j_\varphi , j_\pi , t] &=& \int d^d x \left\{ j_\varphi (x)
{\delta\over\delta j_\pi (x)} + 
j_\pi(x)\Delta {\delta\over\delta j_\varphi (x)}\right.\nonumber \\ 
&+& 
iU\left({\delta\over\delta j_\varphi (x)} + {i\over 2} j_\pi(x)\right)
\left. - iU\left({\delta\over\delta j_\varphi (x)} - {i\over 2}
j_\pi(x)\right)\right\} Z [j_\varphi , j_\pi , t]
\label{twentytwo}
\eea
Its generalization to several scalar fields $\tilde\varphi_a(x)$ is
straightforward -- with $\tilde\pi^2 \rightarrow \tilde\pi_a\tilde\pi_a$,
$\partial_i\tilde\varphi
\partial_i\tilde\varphi\rightarrow\partial_i\tilde\varphi_a\partial_i\tilde\varphi_a$
in $H$ (\ref{twelve}) and commutation relation 
$\left[\tilde\varphi_a(x) , \tilde\pi_b(x^\prime)\right] =
i\delta(x-x^\prime)\delta_{ab}$ we only have to replace
$j_\varphi{\delta\over\delta j_\pi}\rightarrow
j_{\varphi_a}{\delta\over\delta j_{\pi_a}}$, $j_\pi\Delta
{\delta\over\delta
j_\varphi}\rightarrow j_{\pi_a}\Delta {\delta\over\delta j_{\phi_a}}$ in
Eq.
(\ref{twentytwo}). All symmetries of $U(\tilde\varphi)$ are preserved by
the
evolution equation in the sense that an initially symmetric state remains
so
at later time. This does not preclude spontaneous symmetry breaking in the
course of the evolution -- it may be detected by adding a small symmetry
breaking linear term in $U$ or by starting with a slightly asymmetric
initial
state.

Equation (\ref{twentytwo}) is a functional differential equation and its
approximate solution has to proceed by some truncation. For this purpose
it
seems an advantage to switch to the generating functional for the
1PI-Green
functions, $\Gamma[\varphi,\pi,t] = -\ln Z [j,t] + \int d^dx(j_\varphi(x)
\varphi (x) + j_\pi(x)\pi(x))$, where $\varphi(x) = {\partial\ln Z\over
\delta j_\varphi(x)}$, $\pi(x) = {\delta\ln Z\over \delta j_\pi(x)}$. The
derivation of the evolution equation for the effective action $\Gamma$
proceeds as in Ref. \cite{W1} and we only give here the result for the
Hamiltonian (\ref{twenty}) with $\partial_t\Gamma$ a time derivative at
fixed
$\varphi$ and $\pi$
\bea
\partial_t\Gamma &=& -\bigg({\cal L}^{(\Gamma)}_{cl} + \Delta
{\cal L}^{(\Gamma)}_{QM} \bigg) \Gamma \nonumber \\
{\cal L}^{(\Gamma)}_{cl} &=& \int d^dx\bigg\{ \pi(x)
{\delta\over\delta \varphi(x)} + \varphi(x) (\Delta - \mu^2)
{\delta\over\delta \pi(x)} \nonumber \\
&&- \bigg[ \nu ( \varphi^2(x) + G_{\varphi\varphi}(x,x)) + {\lambda\over
2}
\bigg(\varphi^3(x) + 3\varphi(x) G_{\varphi\varphi}(x,x) \nonumber \\
&&-\int
dx_1dx_2dx_3 G_{\varphi\gamma_1}(x,x_1)
G_{\varphi\gamma_2}(x,x_2)\nonumber \\
&& G_{\varphi\gamma_3}(x,x_3) ~~{\delta^3\Gamma\over
\delta\hat\varphi_{\gamma_1} (x_1) \delta\hat\varphi_{\gamma_2}(x_2)
\delta\hat\varphi_{\gamma_3}(x_3)}\bigg) \bigg]
{\delta\over\delta\pi(x)}\bigg\}\nonumber \\
\Delta{\cal L}^{(\Gamma)}_{QM} &=& \int d^dx\bigg\{ {\nu\over
12}~\left({\delta\Gamma\over\delta\pi(x)}\right)^3 + {\lambda\over 8}
\varphi(x) \left({\delta\Gamma\over\delta\pi(x)}\right)^3\bigg\}
\label{twentythree}
\eea
Here 
$G_{\gamma\gamma^\prime}(x,x^\prime) = \langle
\hat{\tilde\varphi}_\gamma(x)
\hat{\tilde\varphi}_{\gamma^\prime}(x^\prime) \rangle_j - 
\hat{\tilde\varphi}_\gamma(x)
\hat{\tilde\varphi}_{\gamma^\prime}(x^\prime)$
is the propagator in presence of sources where $\hat\varphi_\gamma$,
$\gamma
= \varphi , \pi$ is a shorthand for $(\varphi , \pi)$. The propagator can
in turn be expressed by the inverse of the matrix\footnote{Note that
$\Gamma^{(2)}$ has indices
$(x,\gamma)$ and $(x^\prime , \gamma^\prime)$ with $\gamma =
(\varphi,\pi)$.} of second functional derivatives of $\Gamma$, i.e.,
$G_{\gamma\gamma^\prime}(x,x^\prime) =
(\Gamma^{(2)})^{-1}_{\gamma\gamma^\prime} (x,x^\prime)$.
We observe that ${\cal L}^{(\Gamma)}_{cl}$ plays the role of the classical
Liouville operator, with $\varphi^n$ replaced by
$\langle\tilde\varphi^n\rangle$. The difference
$\langle\tilde\varphi^n\rangle -
\varphi^n$ is accounted for by the terms involving $G$. They induce a
dependence of ${\cal L}$ on $\Gamma$ and turn the evolution equation
non-linear. The quantum corrections are all proportional to
$({\delta\Gamma\over\delta\pi})^3$.

The evolution equation (\ref{twentythree}) has a fixed point $\Gamma_*
(\beta)$
corresponding to thermal equilibrium where $\rho = Z^{-1}_0 e^{-\beta H}$,
$Z_0 = {\rm Tr} e^{-\beta H}$ in Eq. (\ref{three}). This can be computed
by
functional integral methods \cite{EQ}. It is far from obvious, however, if
and in what sense the solutions of Eq. (\ref{twentythree}) with
non-equilibrium initial conditions approach this fixed point. A uniform
approach is not possible due to the existence of infinitely many other
fixed
points which correspond to incoherent mixtures of eigenstates of $H$. At
best, the equilibrium fixed point can be approached if we restrict the
discussion to correlation functions of suitably averaged fields (coarse
graining) or to other subsystems. Thermalization can also be achieved by a
coupling to an environment, thus introducing a stochastic element in the
equations of motion. A truncation of $\Gamma [ \varphi, \pi, t]$ may
destroy
the existence of the infinitely many fixed points. Information about
higher
1PI correlations or their precise momentum dependence is omitted in this
way.
It is conceivable that truncated equations have a more uniform approach to
the fixed point than the exact ones. Good truncations should at least
retain
those terms that play an important role in the computation of $\Gamma_*
(\beta)$ by the solution of renormalization group equations in dependence
on
a coarse graining scale \cite{WRG}. A better understanding of the impact
of
truncations will be crucial for the practical use of the present
formalism.

\end{document}